# Computational model for the formation of uniform silver spheres by aggregation of nanosize precursors

**Daniel T. Robb**, **Ionel Halaciuga**, **Vladimir Privman** and **Dan V. Goia**

Center for Advanced Materials Processing, Departments of Chemistry and Physics,
Clarkson University, Potsdam, New York 13699

**Abstract:** We present results of computational modeling of the formation of uniform spherical silver particles prepared by rapid mixing of ascorbic acid and silver-amine complex solutions in the absence of a dispersing agent. Using an accelerated integration scheme to speed up the calculation of particle size distributions in the latter stages, we find that the recently reported experimental results — some of which are summarized here — can be modeled effectively by the two-stage formation mechanism used previously to model the preparation of uniform gold spheres. We treat both the equilibrium concentration of silver atoms and the surface tension of silver precursor nanocrystals as free parameters, and find that the experimental reaction time scale is fit by a narrow region of this two-parameter space. The kinetic parameter required to quantitatively match the final particle size is found to be very close to that used previously in modeling the formation of gold particles, suggesting that similar kinetics governs the aggregation process and providing evidence that the two-stage model of burst nucleation of nanocrystalline precursors followed by their aggregation to form the final colloids can be applied to systems both with and without dispersing agents. The model also reproduced semi-quantitatively the effects of solvent viscosity and temperature on the particle preparation.

**Running Title:** Model for formation of uniform silver spheres

**I. Introduction**

Highly dispersed uniform particles are widely used presently in various areas of technology and medicine and are likely to be incorporated into many other applications in the future. One of the most versatile methods for preparing such systems is by chemical precipitation in homogeneous solutions. While theoretical and computational modeling of chemical synthesis of colloids and nanoparticles has advanced recently, more progress needs to be made in understanding the complex phenomena involved, which take place over multiple length and time scales. In this study, we report computational modeling of the size distribution of secondary polycrystalline colloid particles produced by aggregation of primary nanosize crystalline precursors.

Although initially colloidal dispersions of narrow size distribution were widely thought to consist of monocrystalline particles,[1,2] experimental evidence eventually accumulated[3-13] showing that, in many cases, the final particles are in fact polycrystalline. Since aggregation of diffusing particles generally produces a widening size distribution, this presented a theoretical challenge to explain the narrow size distribution. Several aggregation models were proposed that could produce narrow size distributions.[14-21] In particular, a model was proposed which produced size distributions with narrow (relative) width by a two-stage process, coupling an initial nucleation of primary nanocrystals with the aggregation of these nanocrystals into secondary nanoparticles.[22] In this model, the initial strong nucleation of nanocrystals leads to a sizeable peak of small secondary aggregates. If the nucleation rate of nanocrystals then decreases in time with the proper profile, as occurs naturally in the nucleation process, then the initial peak of secondary aggregates grows to larger sizes, while few additional secondary aggregates are formed. Thus a well-defined peak of secondary particles grows with its average size increasing at least as rapidly as its width, resulting in a narrow size distribution.

An important simplification was to approximately account for the growth of primary particles after nucleation, as well as the absorption of monomers (atoms, molecules) by the secondary particle distribution, with the assumption that the aggregating primary particles were of an (experimentally determined) uniform size throughout the process. In addition, from experimental observations, the secondary particles were assumed to consolidate rapidly into spheres with the bulk density, on time scales much faster than that of the diffusive processes. This allowed the Smoluchowski rate[23] $K_S = 4\pi R_S D_1$ for the diffusive capture of smaller particles with diffusion constant $D_1$ by a larger particle with radius $R_s$, under the assumption of diffusion limited process, i.e., "instantaneous" reaction/merging of the smaller particles with the larger particle, to be used as a basis for the aggregation rates of primary particles.

In general, with $N_s(t)$ the concentration of secondary particles containing $s$ primary particles, the rate equations for aggregation of secondary particles can be written

$$\frac{dN_s}{dt} = \sum_{j=1}^{\lfloor s/2 \rfloor} f_{s-j,j} K_{s-j,j} N_{s-j} N_j - \sum_{i=1}^{\infty} f_{i,s} K_{i,s} N_i N_s , \tag{1}$$

where $\lfloor s/2 \rfloor$ refers to the greatest integer less or equal to $s/2$, and where the first sum is understood to be zero for $s=1$. The Smoluchowski rate for two spherical particles is defined[23,24] as $K_{i,j} = 4\pi (R_i + R_j)(D_i + D_j)$, with, under the instantaneous reaction assumption, the kinetic coefficients $f_{i,j} = 1$ for $i \neq j$, but $f_{i,j} = 1/2$ for $i = j$.

However, it has been observed experimentally and argued theoretically that larger secondary particles almost never merge,[22,24-27] presumably because of the relatively small area of potential contact between their surfaces and kinetic factors, suggesting that $f_{i,j}$ can be taken to be zero for $i, j$ both above a certain size. In addition, the merging of small primary particles (generally ~ 10 nm) does not fully comply with the assumptions behind the Smoluchowski rates both because the concept of a depletion zone becomes suspect in the case of small equally-sized particles, and because one can easily imagine that encounters between such particles may not always results in a merger (violating the instantaneous reaction assumption). Thus, the smallest secondary particles may also require adjusted kinetic factors $f_{i,j}$.

Here we consider experimental data for synthesis of spherical silver colloid particles. The experimental details were reported in Ref. 28, and pertinent results are reproduced here. We then use the model of only singlet (precursor nanocrystal) capture by larger aggregates,[22,24-27] which would correspond to taking $f_{i,j} = 0$ unless $i = 1$ or $j = 1$. This assumption of singlet-dominated aggregation is explained shortly. It was also made in the first application of the two-stage model to the formation of gold nanoparticles.[22] Our aim in this work is to further develop and test the model, as well as establish that the newly studied silver colloid system is described by the same approach used previously for modeling of gold.

In its initial application to the gold system,[22] the model produced a distribution with small relative width, but with an average size 3-5 times too small, indicating that too many secondary particles were produced. A second modeling effort for the gold system[25] demonstrated that reducing the kinetic parameter $f_{1,1}$ to values significantly less than 1, as relaxing the instantaneous reaction assumption for singlet-singlet aggregation would suggest, reduced the number of produced secondary particles and increased the average size of the final secondary particle distribution. We note that the kinetic parameters $f_{1,j}$ for $j>1$ are kept at their diffusion-limited values, 1. An application of the model to the synthesis of CdS particles[26] found that a value of $f_{1,1} = 5.0 \times 10^{-4}$ produced good agreement with the average size of the experimental CdS particle distribution, and that the same value also produced reasonable agreement for the gold system of Refs. 22 and 25.

A second approach was taken with the CdS system,[27] in which the assumption of singlet dominance was relaxed, and a cutoff size $s_{max}$ for secondary particle aggregation was introduced as a parameter, with

$$f_{i,j} = 0, \quad \text{for } i > s_{max} \text{ or } j > s_{max},$$

$$f_{i,j} = \begin{cases} 1, & i \neq j \\ 1/2, & i = j \end{cases}, \quad \text{for } i,j \leq s_{max}.$$

(2)

It was found that a value of the parameter $s_{max} = 25$ produced a similar effect in the CdS system as the parameter $f_{1,1} = 5.0 \times 10^{-4}$ had for the CdS system in Ref. 26. Both adjustments to the coefficients $f_{i,j}$ serve to limit the number of secondary particles produced, and it seems likely that physically realistic modeling of the aggregation process would entail some combination of the two approaches — values of $f_{i,j}$ less than those associated with instantaneous reaction, and a cutoff size above which the merging of two secondary particles is strongly suppressed. Here, in order to more fully explore the parameter space of the model in applying it to the silver system, we have used the less computationally intensive, singlet-dominated formulation. For simplicity, beginning in Sec. II, we will refer to the parameter $f_{1,1}$ as $f$.

The paper is organized as follows. In Sec. II, we summarize the experimental procedure and results reported recently[28] for the production of uniform silver spheres. In Sec. III, we describe the two-stage model for particle growth previously applied to understand the synthesis of gold and cadmium sulfide nanoparticles, and describe the implementation of a computational technique enabling accelerated integration of the two-stage model to longer time scales. In Sec. IV, we revisit the application of the two-stage model to the gold system, using the accelerated integration technique. In Sec. V, we apply the model to silver nanoparticle production, performing a simultaneous fit of the surface tension and equilibrium concentration parameters. We then compare the value of the kinetic parameter required to match the experimental size distribution to the value required in the gold system, and study the effect of solution viscosity and temperature.

**II. Experimental System**

We apply the two-stage model here to chemical synthesis of uniform silver spheres via reduction of silver-polyamine complexes with iso-ascorbic acid. This experimental system was described and characterized in detail recently.[28] The system has commercial relevance as a straightforward way to produce uniform dispersed silver particles that are useful in the production of silver conductive structures in the electronics industry. Several methods, including photoreduction, spray pyrolysis, and precipitation in homogeneous solutions or reverse micelles, are available for the preparation of silver particles. Precipitation in homogeneous solutions remains the most versatile approach, due to the possibility of using a broad range of solvents and a large variety of reducing, dispersing, and complexing agents. Depending on the experimental conditions chosen, the precipitation method investigated in Ref. 28 yields spherical silver particles of a narrow size distribution, with an average diameter ranging from 80 nm to 1.3 μm. From the applications point of view, this method is favorable, because, in contrast to most precipitation protocols, it generates dispersed silver spheres in concentrated systems in the absence of a dispersing agent. As a result, the particles are free of the organic residues that adversely affect their performance in most electronic applications. From the modeling point of view, the dispersing agent should not impact the dynamics of the aggregation of nanosize precursors, because the precursors (as well as the monomers) are below the size range affected by such agents. The results presented in this paper, specifically the similar value of the kinetic parameter, *f*, required to achieve agreement between the computed and experimental average particle size in the gold and silver systems, support this expectation.

As in the previously studied colloidal gold system,[22] here both electron microscopy and X-ray diffraction have shown that the silver spheres were composed of smaller crystalline subunits. The field emission scanning electron microscopy (FESEM) images in Fig. 1, which correspond to the $T = 60$ C case in Ref. 28, reveal the presence of subunits with an approximate size of 12-20 nm. The size of the subunits was also calculated from X-ray diffraction measurements, see Fig. 1(c), using the Scherrer formula.[29] The calculated size of 18-20 nm (as obtained from several diffraction peaks) agreed with the size observed by FESEM, providing strong evidence that the final silver spheres formed by an aggregation process and that the two-stage model should be applicable. The succession of vivid colors during the early stages of the precipitation confirmed the presence of the dispersed subunits and growing aggregates.

A lower magnification electron micrograph of the silver particles produced at $T = 60$ C is shown in Fig. 2a. Their size distribution (Fig. 2b) was obtained by direct measurement of over 200 particles from FESEM images. In Ref. 28, the effects of varying a number of experimental parameters on the reduction process were examined. In Fig. 3, FESEM images of the particles produced at 60 C in both aqueous solvent and in diethylene glycol (DEG), which is considerably more viscous, are shown. In Fig. 4, FESEM images of particles produced at temperatures $T = 40$ C, 60 C, and 80 C are presented. In Sec. V, we will evaluate the extent to which the two-stage model can explain the results in solvents of different viscosity and at varying temperature.

**III. Two-Stage Model of Particle Growth**

The two-stage model for particle growth was developed to explain the narrow size distribution occurring in synthesis of gold nanoparticles.[22,25] The model was motivated by experimental observations[22] that the uniform gold particles were actually polycrystalline, formed by aggregation of many smaller crystalline subunits. As mentioned in the Introduction, a polycrystalline structure in uniform colloids had been observed in many other systems as well.[9-19,26-28] In the model, an initial process of nucleation of clusters (primary particles) is described by standard burst-nucleation theory. The reduction process results in a supersaturated

– 6 –

solution with monomer (here, aqueous neutral metal atom) concentration $c$. Driven by thermal fluctuations, small nanoclusters (embryos) are produced, whose size distribution is controlled by the free energy of an *n*-monomer embryo, of the following form,[22,25,30]

$$\Delta G(n,c) = -nkT \ln(c/c_0) + 4\pi a^2 n^{2/3} \sigma , \qquad (3)$$

where $k$ is the Boltzmann constant, $T$ is the temperature in Kelvin, $c_0$ is the equilibrium concentration of monomers, and $\sigma$ is the effective surface tension. The first term is the free-energy contribution of the "bulk" of the embryo. The second term represents the surface free-energy, proportional to the surface area of the embryo, and therefore to $n^{2/3}$. The effective solute radius $a$, chosen so that the radius of an *n*-solute embryo is $an^{1/3}$, is defined by requiring that $4\pi a^3/3$ is the "unit cell" volume per monomer (including the surrounding void volume) in the bulk material crystal structure.

At small cluster sizes, the surface term dominates, creating a free-energy barrier to the nucleation of a thermodynamically stable cluster. The peak of the nucleation barrier occurs at the critical cluster size,

$$n_c = \left[ \frac{8\pi a^2 \sigma}{3kT \ln(c/c_0)} \right]^3 . \qquad (4)$$

For $n < n_c$, the concentration $P(n,t)$ of *n*-monomer nanoclusters is assumed to follow a thermal distribution,

$$P(n,t) = c \exp\left[ \frac{-\Delta G(n,c)}{kT} \right] . \qquad (5)$$

The rate of production of supercritical clusters is then given[22] by

$$\rho(t) = K_{n_c} c P(n_c,t) = K_{n_c} c^2 \exp\left[ \frac{-\Delta G(n_c,c)}{kT} \right] , \qquad (6)$$

where

$$K_n = 4\pi \left(a + an^{1/3}\right)\left(D_{\text{atom}} + D_{\text{atom}} n^{-1/3}\right) \approx 4\pi a n^{1/3} D_{\text{atom}} \qquad (7)$$

is the rate constant in the Smoluchowski description of diffusive capture of particles.[23] These expressions involve various approximations described in earlier works.[22,25,30] A key quantity is the rate at which monomers are consumed by the growing supercritical clusters, thereby reducing the supply of monomers available to constitute the thermal distribution Eq. (5),

$$\frac{dc}{dt} = -n_c \rho(t) \ . \tag{8}$$

The nucleation rate $\rho(t)$ from Eq. (6) is then used as input to a kinetic model for the aggregation of the primary particles into secondary, polycrystalline aggregates. The master equation for the aggregation process under the assumption of singlet-dominance takes the form

$$\frac{dN_s}{dt} = K_{s-1} N_1 N_{s-1} - K_s N_1 N_s \ , \quad \text{for } s > 2 \ , \tag{9}$$

where $N_s(t)$ is the time-dependent number density (per unit volume) of secondary particles consisting of $s$ primary particles (the expressions for $s = 1, 2$ will be given shortly). The attachment rate, $K_s$, is modeled using the Smoluchowski rate for diffusive capture, already encountered in the Introduction, of the form

$$K_s = 4\pi \left( R_1 + R_s \right)(D_1 + D_s) \ . \tag{10}$$

Here $R_s = 1.2 r s^{1/3}$ is the radius of a secondary particle containing $s$ primary particles, with the factor 1.2 calculated as $(0.58)^{-1/3} \simeq 1.2$, where 0.58 is the typical filling factor for random loose packing of spheres.[31] The diffusion constant of a secondary particle containing $s$ primary particles is given by $R_s = D_1 s^{-1/3}$. The average radius $r$ of the primary particles is available experimentally (from X-ray diffraction), from which the primary particle diffusion constant can be calculated using the Stokes-Einstein relation: $D_1 = kT/6\pi\eta r$. Note that in Ref. 22, the attachment rate was approximated as $K_s = 4\pi R_s D_1$, which corresponds to the limit $s \gg 1$. Here we follow Ref. 25 and use the full expression Eq. (10).

For the case $s = 2$, the master equation was taken in Ref. 25 to be

$$\frac{dN_2}{dt} = fK_1 N_1^2 - K_2 N_1 N_2 \ . \tag{11}$$

As discussed in the Introduction, for the assumption of instantaneous reaction (diffusion-limited kinetics), the kinetic parameter $f$ would be 1/2. Here, using the singlet-dominated approach of Eq. (9), we will eventually use $f \ll 1/2$ to limit the number of secondary particles produced. The evolution of the singlet ($s = 1$) population is given by

$$\frac{dN_1}{dt} = \rho(t) - \sum_{j=2}^{\infty} j \frac{dN_j}{dt} . \tag{12}$$

To make the model numerically tractable, the diffusive growth of already-nucleated primary particles is not explicitly included, but is incorporated by the use of the experimentally determined primary particle size. In combination with Eq. (8), this can lead to the conservation of matter being violated. Thus, the final resulting distribution $N_s(t)$ is regarded as relative,[22,25] and must be normalized to correspond to the total amount of matter initially present, given by the initial monomer concentration, $c(t=0)$.

**IV. Simulation Results for Gold**

In Ref. 26, it was found that kinetic parameter $f = 5.0 \times 10^{-4}$ produced approximate agreement with the experimental average particle size at saturation for both CdS and Au systems. Here, we will compare the present Ag system with the Au system of Refs. 22 and 25. With the benefit of the accelerated integration scheme (detailed in Appendix A) to access longer time scales, we adopt an explicit criterion for saturation. In this section, we thus re-examine the Au system in our more explicit numerical framework, to enable a direct comparison with the Ag system.

We first illustrate what we mean by the "saturation" of the secondary particle distribution. In Fig. 5, the secondary particle distribution is shown at times $t = 0.1, 1.0, 5.0$, and 10.0 s, for parameter values $c_0 = 1.0 \times 10^{15} \, \text{m}^{-3}$ (taken from the literature[32] in Ref. 22) and $\sigma = 0.555 \, \text{N/m}$. The distribution moves outward toward larger sizes between $t = 0.1$ s and $t = 5$ s, and then begins to move more slowly between $t = 5$ s and $t = 10$ s. With the supply of primary particles to feed the secondary distribution nearly exhausted, the distributions at times $t > 10$ s (not shown) are found to lie almost on top of the distribution at $t = 10$ s. It is in this sense that the distribution is said to saturate at a time scale of approximately 10 s. (We note that in Ref. 25, because a rough criterion for the saturation time was used, a somewhat different value of $\sigma = 0.51 \, \text{N/m}$ was identified as best fitting the experimental time scale.)

The distributions in Fig. 5, as well as in our other calculations presented later, have been normalized so that the total amount of matter, including monomers, primary particles, and the secondary distribution, is constant in time and equal to the initial monomer concentration $c(t=0)$. The right vertical axis is given in units of concentration per unit diameter ($m^{-3}/m = m^{-4}$), so that the integral of the distribution $\int_0^\infty N(D)\,dD = c(t=0)$ has the correct units of concentration ($m^{-3}$). The left vertical axis expresses the total concentration, over a diameter interval of 1 nm, in the more commonly used units of mol/L.

Here we have adopted the following numerical criterion for the saturation time. First, we use a highly accelerated integration scheme, based on the treatment in Ref. 33, to approximate the average size of the particle distribution as a function of time. These equations are presented at the end of Appendix A. The average size at full saturation, $d_{full}$, is recorded at the first time, $t_{full}$, for which the average size at earlier time $t_{full}/2$ is greater than $0.995\,d_{full}$. We then integrate the system a second time, using the accelerated scheme presented in the first part of Appendix A. This scheme is slightly more accurate and still much faster than using the full system of Eqs. (6)-(12), and furthermore, it yields the actual size distributions. The simulated time $t_{sat}$, to be compared to the time of the end of observable color changes of the solution in experiment, is then set to when the average size reaches 90% of the value at full saturation, i.e., $d(t_{sat}) = 0.9 d_{full}$. Other criteria for correspondence with the experimentally observed saturation time are possible, but would not change our main conclusions.

Using the criterion for saturation described above, the saturation time was found to be $t_{sat} = 7.5$ s, in approximate agreement with the experimentally observed saturation time of $\sim 10$ s. With kinetic factor $f = 1/2$, the average particle size at full saturation was $d_{full} = 0.624 \pm 0.02~\mu m$, as compared to the experimentally observed final size $d_{expt} = 2.0 \pm 0.2~\mu m$. Using $f = 10^{-4}$, the saturation time was unchanged at $t_{sat} = 7.5$ s, but the diameter increased to $d_{sat} = 1.973 \pm 0.035~\mu m$, in close agreement with experiment.

**V. Simulation Results for Silver**

In applying the two-stage model to the experimental Ag system described in Sec. II, we found that the available literature values[34] for equilibrium concentrations of Ag (in highly purified, neutral, degassed water) were in the range $c_0 = 1.4\text{-}2.0 \times 10^{20}~m^{-3}$. Given the large difference from the value for gold quoted above ($c_0 \simeq 1.0 \times 10^{15}~m^{-3}$), as well as the evidence for surface contamination by surface oxides, adsorbed $O_2$, and/or dissolved $O_2$, in the experiments for silver cited in Ref. 34, we decided to treat not only the surface tension but also the equilibrium concentration as a free parameter. A series of simulations were run at different values of the two parameters ($\sigma, c_0$), using kinetic parameter $f = 0.5$. The following parameters were used, appropriate for the $T = 60 C$ case in aqueous solvent described in Sec. II: $c(t = 0) = 5.576 \times 10^{25}$ m$^{-3}$, $T = 60 C = 333 K$, $a = 1.60 \times 10^{-10}$ m, $r_1 = 8.7 \times 10^{-9}$ m, $D_{atom} = 3.63 \times 10^{-9}$ m$^2$s$^{-1}$, $D_1 = 5.97 \times 10^{-11}$ m$^2$s$^{-1}$. The diffusion constants $D_{atom}$ and $D_1$ were calculated using the Stokes-Einstein relation with viscosity $\eta_{H_2O} = 0.4666$ mPa s at 60 C, and with the approximate hydrodynamic radii $r_{atom} = 1.44 \times 10^{-10}$ m and $r_1 = 8.7 \times 10^{-9}$ m, respectively.

The results are illustrated in Fig. 6, which combines a contour plot of the saturation time with a representation of the average particle sizes (at full saturation) on a regular grid of points ($\sigma, c_0$). Interestingly, the saturation time — determined using the criterion $d(t_{sat}) = 0.9 d_{full}$ described above — depends in a systematic way on both the surface tension and the equilibrium concentration. It varies over an enormous range, from $10^{-5}$ s in the lower left corner of the figure to $10^{13}$ s in the upper right corner. We caution, however, that the assumption of the dominance of the singlet-cluster attachment process becomes suspect for saturation times above ~ 300-500 s, at which point the processes of cluster-cluster attachment and Ostwald ripening become relevant. Thus the saturation time of ~ $10^{13}$ s for parameter values $\sigma = 0.55$ N/m, $c_0 = 10^{20}$ m$^{-3}$, while accurate for the mathematical model, would likely be much lower in experiment.

The diameter $d_{full}$ at full saturation also varies widely, from 70.7 nm at $\sigma = 0.40$ N/m, $c_0 = 10^{17}$ m$^{-3}$, to 17.8 $\mu$m at $\sigma = 0.55$ N/m, $c_0 = 10^{20}$ m$^{-3}$. For the particles shown in Fig. 2, the experimentally observed time scale for saturation was ~ 100 s. This time scale can be reproduced in simulation by using values of ($\sigma, c_0$) drawn from anywhere along the $t = 100$ s contour in Fig. 6. A previous study[35] of the surface tension of silver nanoparticles reported the surface tension in colloidal solution as $\sigma = 0.525$ N/m. We therefore provisionally identify the joint parameter values $\sigma = 0.525$ N/m, $c_0 = 1.15 \times 10^{18}$ m$^{-3}$, which results in a saturation time $t_{sat} = 96.7$ s, as the best fit to our experimental results. However, it is important to note that the value $\sigma = 0.525$ N/m is at the lower end of a wide range of values for the surface tension of silver particles reported in the literature.[35] Our main finding at this point is not that the parameter values $\sigma = 0.525$ N/m, $c_0 = 1.15 \times 10^{18}$ m$^{-3}$ are precisely correct, but that the joint values of ($\sigma, c_0$) must be drawn from the relatively narrow $t = 100$ s contour region.

The particle size distribution as a function of time, for the best-fit parameter values $\sigma = 0.525$ N/m, $c_0 = 1.15 \times 10^{18}$ m$^{-3}$, and kinetic parameter $f = 0.5$, is shown in Fig. 7 (the left data series). The average size at full saturation is $d_{full} = 0.349 \pm 0.008$ $\mu$m, which is smaller than the experimentally observed particle size $d_{expt} = 0.967$ $\mu$m. Using kinetic parameter $f = 5.0 \times 10^{-4}$ leaves the saturation time unchanged, but increases the average size at full saturation to $d_{full} = 0.970 \pm 0.003$ $\mu$m, in quantitative agreement with the experimental results. Thus, the value of the kinetic parameter, $f = 5.0 \times 10^{-4}$, required to match the experimentally observed final particle size in this silver system is quite close to the value $f = 1.0 \times 10^{-4}$ which was required in Sec. II to match the experimental size in the gold system. In addition, the value is equal to that found in Ref. 26 to be required to match the experimental particle size in the CdS and Au systems. This suggests that the kinetic parameter represents similar physics at work in the three systems.

We next investigate the degree to which the two-stage model reproduces the effects of varying the solvent used in synthesizing the Ag particles. As illustrated in Fig. 3, the use of a more viscous solvent than water, namely diethylene glycol (DEG), with[36] $\eta_{DEG} = 7.64$ mPa s at $T = 60$ C, resulted in a much smaller final particle size. The saturation time was observed to be quite close to the aqueous system (~ 100 s). Using several reasonable assumptions, we can account semi-quantitatively for this significant decrease in the final particle size. First, we assume that the diffusion constants $D_{atom}$ and $D_1$ of both individual silver atoms and primary particles in DEG are related to those in water by the hydrodynamic relation $D_{DEG} = (\eta_{H_2O}/\eta_{DEG}) D_{H_2O}$. To achieve a saturation time ~ 100 s, it is then necessary to decrease slightly the value(s) of $\sigma$ and/or $c_0$, which is consistent with the fact that alcohols are usually found to have lower equilibrium concentrations than water. We chose the values $\sigma = 0.505\,\text{N/m}$ and $c_0 = 1.0 \times 10^{18}\,\text{m}^{-3}$.

As shown in Fig. 8, using kinetic parameter $f = 5 \times 10^{-4}$ resulted in an average size at full saturation $d_{\text{full}} = 0.693\,\mu\text{m}$ and a saturation time $t_{\text{sat}} = 77.1\,\text{s}$. However, it is reasonable to suppose that the kinetic coefficient $f$ in DEG is somewhat larger than that in water, since the increased rotational viscosity of DEG may result in a higher probability of an encounter between singlets (primary particles) producing a permanent merger of the two particles — indeed, rotational diffusion was identified as a contributing factor to the suppression of the formation of dimers in protein crystallization in solutions.[37] As illustrated in Fig. 8, increasing the value of $f$ to 0.005 or 0.05 brings the average size at saturation closer to the experimental value of 80 nm. Still, the remaining factor of 3-4 indicates that our model does not completely capture the behavior in the DEG solvent. Resolving this discrepancy may require a more accurate understanding of the kinetic coefficients $f_{i,j}$ in Eq. (1).

Finally, we investigate the effect of the temperature on the final size distribution, attempting to model the experimental results shown in Fig. 4. Table I shows the saturation times and the average sizes at saturation at the three temperatures $T$ = 40, 60, and 80 C, as well as the saturation time and average size (at full saturation) in simulation, using the best-fit parameters $\sigma = 0.525\,\text{N/m}$, $c_0 = 1.15 \times 10^{18}\,\text{m}^{-3}$, and kinetic parameter $f = 5.0 \times 10^{-4}$. The calculated average diameters are in reasonable agreement with those from experiment. The saturation times show the same trend as in the experiment, decreasing from $T$ = 40 C to $T$ = 80 C, but vary over a much larger range in simulation than in experiment. As noted above, the simulated saturation time $t_{\text{sat}} = 12530\,\text{s}$ at $T$ = 40 C would be expected to be shorter in experiment, due to competition from other kinetic processes of ripening and broadening past ~ 300-500 s. The discrepancy in the predicted time scale at $T$ = 80 C is more difficult to understand, but may result from the (slight) dependence of both $\sigma$ and $c_0$ on temperature. As seen in Fig. 6, even small changes in these parameters can significantly affect the time scale. The actual size distributions at each temperature are shown in Fig. 9.

In conclusion, we have used the two-stage model previously applied to Au and CdS systems to model the production of uniform spherical silver particles. The use of an accelerated integration scheme applicable to later times enabled the exploration of the parameter space of both the surface tension $c_0$ and equilibrium concentration $c_0$, with the conclusion that they play a comparable role in determining the saturation time for particle production. We determined that the kinetic parameter $f$ in the singlet-dominated aggregation scheme does not noticeably affect the saturation time in the model, and that the value of $f$ required to match the average final particle size in the Ag system is very similar to that required in the Au and CdS systems. This provides further evidence that the model captures important aspects of the kinetics of aggregation in chemical synthesis across different experimental systems. With parameters optimized for $T = 60$ C in aqueous solution, the model was found to account semi-quantitatively for experimental results at different temperatures and in the solvent diethylene glycol (DEG). It seems likely that a more accurate model for the kinetics of aggregation at small primary particles size, intermediate between the singlet-dominated approach with $f \ll 1/2$ and the approach of Ref. 27 involving a cutoff size for primary particle aggregation, may be required to produce quantitative agreement with experiment over a range of temperatures and solvents.

The authors gratefully acknowledge instructive discussions with E. Matijević and I. Sevonkaev. This research was supported by the NSF under grant DMR-0509104.

**Appendix A. Accelerated Integration Schemes for the Two-Stage Model**

The numerical integration of Eqs. (6)-(12) slows down considerably at later times, as the size of the largest secondary particle, and therefore the size of the array storing the distribution $N_s(t)$, increases. Moreover, as illustrated in Sec. 4, the size distribution approaches its saturated form very slowly, in "logarithmic" time. These two factors combine to make the total computational time needed to reach saturation, using the full system of Eqs. (6)-(12), extremely long. Large-scale numerical simulations of Ref. 26 thus used a complicated adaptive numerical rediscretization scheme (in both the time and cluster-size variables).

Fortunately, it is possible to perform a useful approximate integration of the system. Once the loss of singlet secondary particles due to production of new doublet secondary particles becomes significantly smaller than the loss due to absorption by the entire secondary distribution, one can assume[33] that no new doublet particles enter the secondary particle distribution. The other key assumption[33] is to write Eq. (9) in continuous form, and to neglect the "diffusive" second-derivative term as well as higher-order terms. While this approximation scheme underestimates the width of the evolving distribution, the average size of the distribution can be approximated very well, and the scheme is quite efficient.

Combining Eqs. (11) and (12), one finds

$$\frac{dN_1}{dt} = \rho(t) - 2fK_1 N_1^2 - N_1 \sum_{j=2}^{\infty} K_j N_j \ . \tag{A1}$$

The system of Eqs. (6)-(12) can typically be integrated fairly quickly to a time $t = t_1$ where new doublet production is insignificant, i.e., where $\Xi \equiv 2fK_1 N_1 \big/ \sum_{j=2}^{\infty} K_j N_j \ll 1$ in Eq. (A1). For example, for the simulation at $f = 0.5$ shown in Fig. 7, with time step $\Delta t = 2.0 \times 10^{-6}$ s, we found $\Xi \approx 3 \times 10^{-4}$ at $t = 2.0$ s. The integration to $t = 2.0$ s required less than 10 min on a 1.6 GHz Intel Core 2 Duo processor. For the other simulation shown in Fig. 7, at $f = 5.0 \times 10^{-4}$, a similar value $\Xi \approx 7.7 \times 10^{-5}$ was reached at $t = 0.2$ s, in less than 5 min of computation time.

In the development of the approximation scheme,[33] it is then suggested to assume that $\Xi = 0$ by setting $K_1 = 0$. However, we have found numerically that there is a delicate numerical balance on the r.h.s. of Eq. (A1), with

$$\rho(t) - N_1 \sum_{j=2}^{\infty} K_j N_j \approx 2fK_1 N_1^2 , \tag{A2}$$

and furthermore with $\dfrac{dN_1}{dt} \ll 2fK_1 N_1^2$. Thus it is best to maintain the term $2fK_1 N_1$ in Eq. (A1) to accurately determine the evolution of the singlet concentration $N_1$. The assumption that doublet production has stopped is still explicitly applied, by propagating the size distribution (starting from the chosen time $t = t_1$, where $\Xi \ll 1$) to larger sizes, without introducing any new secondary particles into the shoulder of the distribution. This propagation is modeled by a continuous version of Eq. (9), with only the first-order ("drift") term retained,

$$\frac{dN(s,t)}{dt} = -N_1(t) \frac{\partial}{\partial s}\left[ K(s) N(s,t) \right] . \tag{A3}$$

Eq. (A1) becomes

$$\frac{dN_1}{dt} = \rho(t) - 2fK_1 N_1^2 - N_1(t) \int_0^{\infty} ds \left[ K(s) N(s,t) \right] . \tag{A4}$$

The solution to Eq. (10) can be written in terms of the auxiliary variable,[33]

$$\tau(t) = \int_0^t dt' N_1(t') . \tag{A5}$$

One also introduces the associated function $u(s,\tau)$, defined by the relation

$$\tau(t) = \int_{u(s,\tau(t))}^{s} \frac{ds'}{K(s')} . \tag{A6}$$

Eq. (A6) implies the following relationship between the differentials,

$$d\tau = \frac{ds}{K(s)} - \frac{du}{K(u)} . \tag{A7}$$

Using Eqs. (A5) and (A7), one can verify after some algebra that Eq. (A3) is solved by

$$N(s,t) = \frac{K(u(s,\tau(t)))}{K(s)} N(u(s,\tau(t)),0) , \tag{A8}$$

which relates the secondary particle distribution at time $t$ to the distribution at time $t = 0$, i.e., the distribution at the final time $t = t_1$ of the integration of the full set of Eqs. (6)-(12). The function $u(s,\tau(t))$ corresponds to the initial position of the secondary particles located at position $s$ at later time $t$. One can also define the inverse function $s(u,\tau(t))$ implicitly from Eq. (A6). Since the initial distribution starts at $u = 0$, the distribution $N(s,t)$ of Eq. (A8) starts at $s_{\min}(u,\tau(t))$, defined by

$$\tau = \int_0^{s_{\min}(u,\tau(t))} \frac{ds'}{K(s')} \ . \tag{A9}$$

The integral on the r.h.s. of Eq. (A4),

$$F(\tau(t)) \equiv \int_0^\infty ds \left[ K(s) N(s,t) \right] , \tag{A10}$$

can be re-expressed in the form

$$F(\tau(t)) \equiv \int_0^\infty du \left[ K(s(u,\tau(t))) N(u,0) \right] . \tag{A11}$$

The distribution $s(u,\tau(t))$ at a fixed value of $u$ can be propagated forward using Eq. (A7), which in this case becomes

$$ds = K(s(u,\tau(t))) d\tau = K(s(u,\tau(t))) N_1(t) dt \ . \tag{A12}$$

The integration of Eq. (A4) must be carried out with a sufficiently short time step $\Delta t$, in order to avoid a numerical instability associated with the delicate balance of terms on its r.h.s. The required size for the time step depends on the model parameters — once the time step is lowered enough to avoid the numerical instability, no accuracy is gained by reducing it further. However, we found that for the form of $K(s)$ used in the two-stage model, cf. Eq. (10), Eq. (A12) could be integrated with a longer time step $\Delta t'$, because of the slow change of $K(s) \sim s^{1/3}$ and $N_1(t)$ with respect to $s$ and $t$, respectively. We determined empirically that no significant increase in accuracy was achieved by taking $\Delta t'$ lower than $2000\Delta t$.

The accelerated integration scheme can be summarized as follows:

1. Integrate the full system Eqs. (6)-(12) up to a time $t = t_1$ at which the ratio $\Xi = 2 f K_1 N_1 \Big/ \sum_{j=2}^\infty K_j N_j \ll 1$. We found $\Xi \approx 10^{-4}$ to give sufficient accuracy.

2. Use the final values $c(t_1)$, $N_1(t_1)$, and $N_u(t_1)$ from step 1 as initial conditions in the following set of equations:

– 17 –

$$\frac{dc}{dt} = -n_c \rho(t) , \qquad \frac{dN_1}{dt} = \rho(t) - 2fK_1 N_1^2 - N_1(t)F(\tau(t)) ,$$

(A13)

$$\frac{d\tau}{dt} = N_1 , \qquad \left.\frac{ds(u,\tau(t))}{dt}\right|_u = K(s(u,\tau(t)))N_1(t) .$$

The variables $n_c$, $\rho$, and $F(\tau(t))$ are defined in Eqs. (4), (6), and (A11), respectively.

3. Perform the integration of $c$, $N_1$, and $\tau$ using a sufficiently small time step $\Delta t$, as discussed above. However, do the integration of $s(u,\tau(t))$, for each $u$ in the initial distribution $N(u,0)$, using a much longer time step $\Delta t'$, e.g., $\Delta t' = 2000\, \Delta t$.

4. At any desired time, calculate the evolved distribution $N(s,t)$ using Eq. (A8).

If one only needs the average size and width of the evolving distribution under the above approximation, and not the actual distribution $N(s(u,\tau(t)))$, it is possible to estimate these even more efficiently, with only slightly less accuracy.[33] We have used the method described below to estimate the "fully saturated" size of the distribution required for our criterion for saturation.

The expression for the average size, found by differentiating

$$\langle s \rangle_t = \frac{\sum_{j=2}^{\infty} j\, N(j,t)}{\sum_{j=2}^{\infty} N(j,t)} \equiv \frac{1}{M}\sum_{j=2}^{\infty} j\, N(j,t) \tag{A14}$$

applying Eq. (12) (conservation of matter), and integrating, is

$$\langle s \rangle_t = \langle s \rangle_0 + \frac{1}{M}\left[ N_1(0) - N_1(t) + \int_0^t \rho(t')dt' \right] . \tag{A15}$$

After the initial integration with the full set of Eqs. (6)-(12) up to $t = t_1$, we found empirically that $N_1(t) \ll \int_0^t \rho(t')dt'$, a condition fulfilled even more strongly at later times. Thus we can neglect $N_1(t)$ in Eq. (A15), and estimate $\langle s \rangle_t$ simply by finding $\rho(t)$ via Eq. (8).

The standard deviation of the distribution *in the accelerated scheme*, i.e., ignoring further production of doublets and neglecting second- and higher-order terms in the continuous version of Eq. (9), can be estimated with the use of Eq. (A8). We define the width $W(t)$, an approximation to the standard deviation, by the relation

$$M\langle s\rangle_t = N(\langle s\rangle_t, t)\langle s\rangle_t W(t), \tag{A17}$$

which leads directly to

$$W(t) = \frac{N(\langle s\rangle_0, 0)}{N(\langle s\rangle_t, t)} W(0). \tag{A18}$$

Making the approximation $\langle s\rangle_0 \approx u(\langle s\rangle_t, \tau(t))$, we can apply Eq. (A8) to Eq. (A18) to find

$$W(t) \approx \frac{K(\langle s\rangle_t)}{K(\langle s\rangle_0)} W(0) > W(0). \tag{A19}$$

We emphasize again that the width estimated by Eq. (A19) is systematically lower than that produced in the full integration of Eqs. (6)-(12), which includes second-order and higher "diffusive" terms.

## References

1. V. K. LaMer and R. J. Dinegar, *J. Am. Chem. Soc.* **72**, 4847 (1950).
2. V. K. LaMer, *Ind. Eng. Chem.* **44**, 1270 (1952).
3. D. Murphy-Wilhelmy and E. Matijević, *J. Chem. Soc. Faraday Trans.* **80**, 563 (1984).
4. E. Matijević and D. Murphy-Wilhelmy, *J. Colloid Interface Sci.* **86**, 476 (1982).
5. E. Matijević and P. Scheiner, *J. Colloid Interface Sci.* **63**, 509 (1978).
6. U. P. Hsu, L. Ronnquist, and E. Matijević, *Langmuir* **4**, 31 (1988)
7. M. Ocana and E. Matijević, *J. Mater. Res.* **5**, 1083 (1990).
8. M. Ocana, C. J. Serna and E. Matijević, *Colloid Polym. Sci.* **273**, 681 (1995).
9. S. H. Lee, Y. S. Her and E. Matijević, *J. Colloid Interface Sci.* **186**, 193 (1997).
10. L. H. Edelson and A. M. Glaeser, *J. Am. Chem. Soc.* **71,** 225 (1988).
11. J. K. Bailey, C. J. Brinker and M. L. Mecartney, *J. Colloid Interface Sci.* **157**, 1(1993).
12. M. P. Morales and T. Gonzales-Carreno, C. J. Serna, *J. Mater. Res.* **7**, 2538 (1992).
13. M. Ocana, M. P. Morales and C. J. Serna, *J. Colloid Interface Sci.* **171**, 85(1995).
14. W. T. Scott, *J. Atmos. Sci.* **25**, 54 (1968).
15. K. Higashitani, *J. Chem. Eng. Jpn.* **12**, 460 (1979).
16. J. Th. G. Overbeek, *Adv. Colloid Interface Sci.* **15**, 251 (1982).
17. R. C. Flagan, *Ceramic Trans.* **1**(A), 229 (1988).
18. J. A. Dirksen and T. A. Ring, *Chem. Eng. Sci.* **46**, 2389 (1991).
19. G. H. Bogush and C. F. Zukoski, *J. Colloid Interface Sci.* **142**, 1 (1991).
20. A. van Blaaderen, J. van Geest and A. Vrij, *J. Colloid Interface Sci.* **154**, 481 (1992).
21. F.-P. Ludwig and J. Schmelzer, *J. Colloid Interface Sci.* **181**, 503 (1996).
22. V. Privman, D. V. Goia, J. Park and E. Matijević, *J. Colloid Interface Sci.* **213**, 36 (1999).
23. R. v. Smoluchowski, *Z. Phys. Chem.* **29**, 129 (1917).
24. V. Privman, *Mechanisms of Diffusional Nucleation of Nanocrystals and Their Self-Assembly into Uniform Colloids*, in: *Proc. Conf. ITP 2007*, Pages 17-1 to 17-15, edited by S. S. Sadhal (Bulgarian Nat. Academy Press, Sofia, 2007).
25. J. Park, V. Privman and E. Matijević, *J. Phys. Chem. B* **105**, 11630 (2001).
26. S. Libert, V. Gorshkov, V. Privman, D. Goia and E. Matijević, *Adv. Colloid Interface Sci.* **100-102**, 169 (2003).
27. S. Libert, V. Gorshkov, D. Goia, E. Matijević and V. Privman, *Langmuir* **19**, 10679 (2003).

28. I. Halaciuga and D. V. Goia, *J. Mater. Res.* **23**, 1776 (2008).
29. A. L. Patterson, *Phys. Rev.* **56**, 978 (1939).
30. D. T. Robb and V. Privman, Langmuir **24**, 26 (2008).
31. R. German, *Particle Packing Characteristics* (Metal Powder Industries Federation, Princeton, 1989).
32. W. F. Linke, *Solubilities of Inorganic and Metal–Organic Compounds*, 4th edition, Vol. **1**, p. 243 (Van Hostrand, Princeton, 1958).
33. V. Privman, *Mater. Res. Soc. Symp. Proc*. **703**, Article T3.3, 577 (2002).
34. D. E. Gray, *American Institute of Physics Handbook,* 3rd edition (McGraw-Hill, New York, 1972).
35. S. F. Chernov, Y. V. Fedorov, V. N. Zakharov, *J. Phys. Chem. Solids* **54**, 963 (1993).
36. T. Sun and A. S. Teja, *J. Chem. Eng. Data* **48**, 198 (2003).
37. C. N. Nanev, *Crystal Growth and Design* **7**, 1533 (2007).

**Table 1.** Experimental and simulated saturation times and average diameter at saturation for the silver system at temperatures $T$ = 40, 60, and 80 C. For the simulation, the best-fit values $\sigma = 0.525$ N/m, $c_0 = 1.15 \times 10^{18}$ m$^{-3}$ and kinetic parameter $f = 5.0 \times 10^{-4}$ were used.

|  | Experiment | | Simulation | |
| --- | --- | --- | --- | --- |
| $T$ ( C) | $t_{sat}$ (s) | $d_{full}$ (nm) | $t_{sat}$ (s) | $d_{full}$ (nm) |
| 40 | 320 | 1490 | 12530 | 1730 |
| 60 | 100 | 967 | 96.7 | 970 |
| 80 | 30 | 475 | 1.4 | 699 |

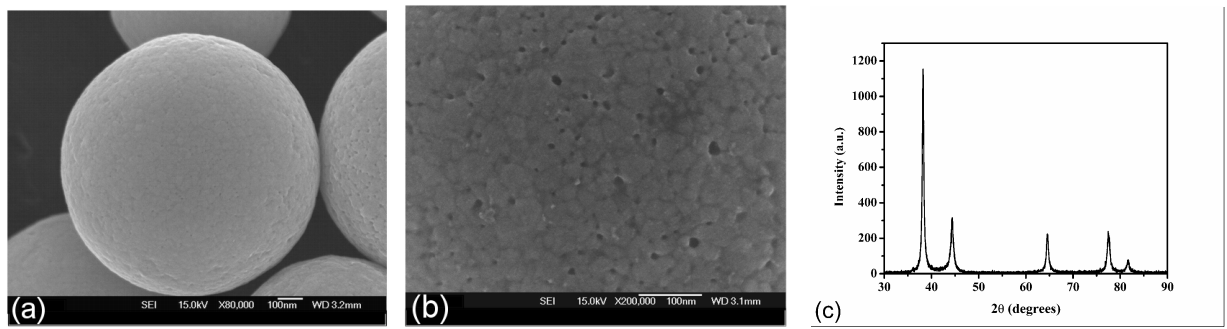

**Figure 1. (a)** Field emission scanning electron micrographs (FESEM) of final silver particles obtained by rapidly adding 60 cm$^3$ of an aqueous solution of iso-ascorbic acid (0.44 mol dm$^{-3}$) to 440 cm$^3$ of an aqueous solution of Ag-Ethylenediamine complex (0.1 mol dm$^{-3}$) at 60 C. **(b)** Higher magnification FESEM image of silver particle surface, suggesting the presence of subunits of diameter ~ 12-20 nm. **(c)** Diffraction peaks measured from X-ray powder diffraction analysis of silver particles.

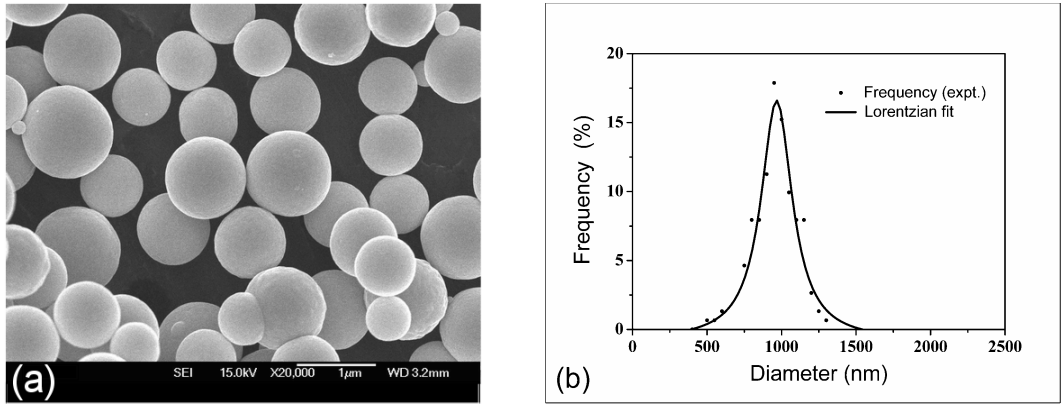

**Figure 2.** **(a)** FESEM image of spherical silver particles produced at 60 C by the method of Ref. 28. **(b)** Histogram of particle diameters taken from the FESEM images. The Lorentzian fit[28] to the histogram had mean value $967 \pm 6$ nm and width $255 \pm 23$ nm.

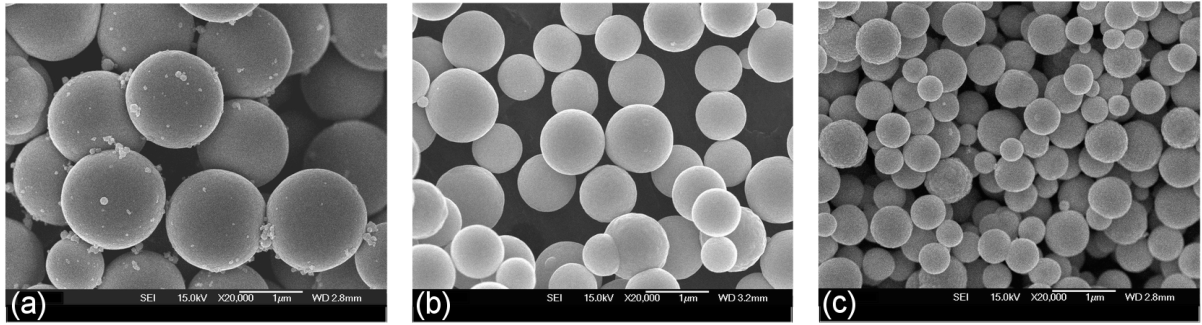

**Figure 3.** FESEM images of silver particles produced at temperatures **(a)** 40 C, **(b)** 60 C, and **(c)** 80 C. Each of the images is shown on the same scale.

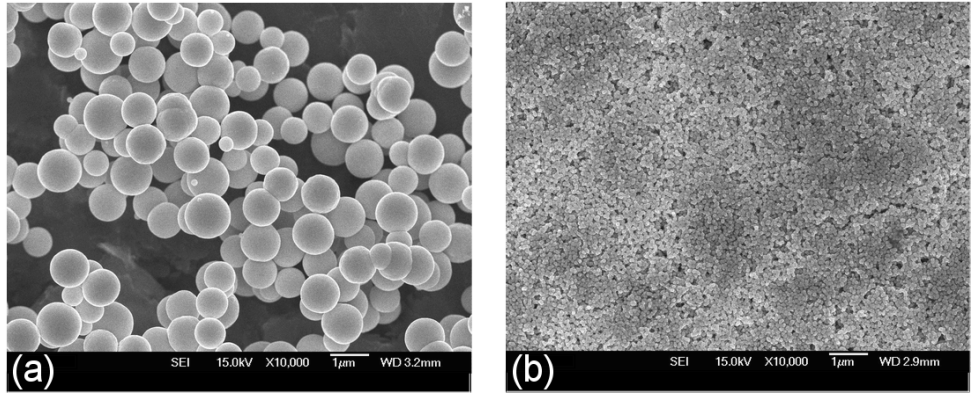

**Figure 4.** FESEM images of silver particles produced at $T = 60$ C, in **(a)** aqueous solvent and **(b)** diethylene glycol (DEG).

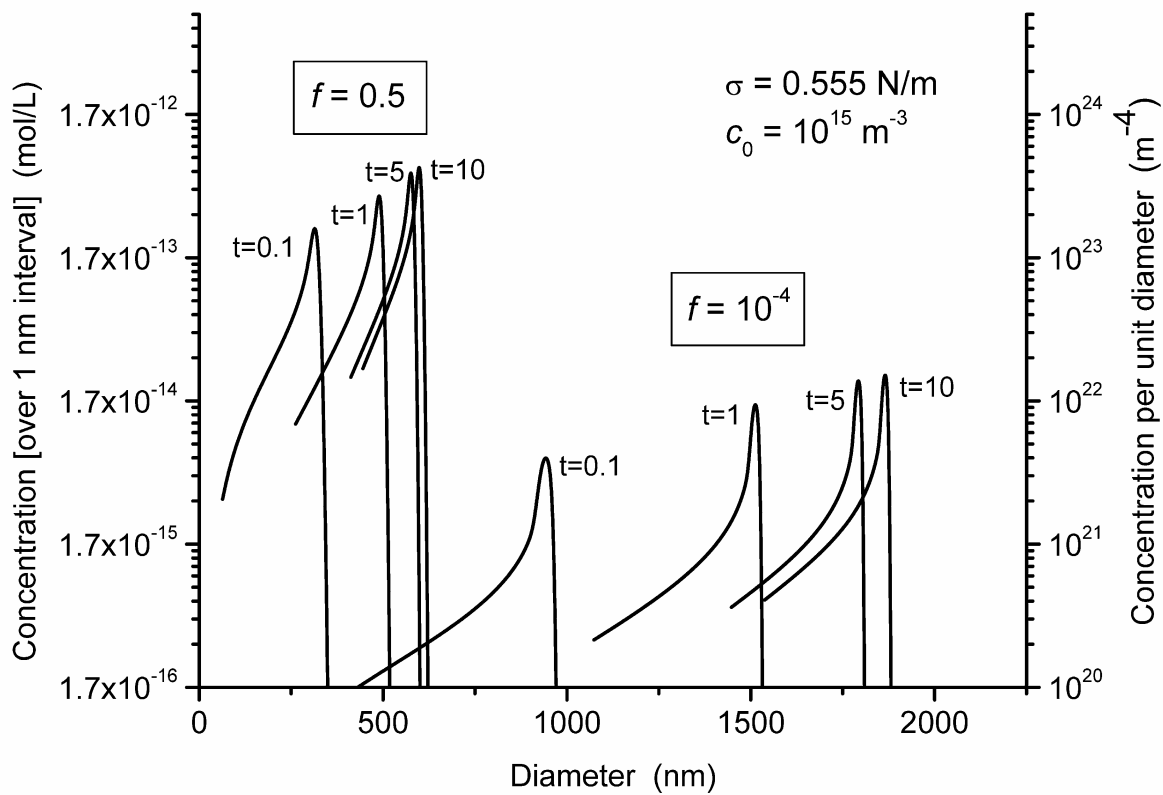

**Figure 5.** Secondary particle size distributions for two-stage model of particle growth applied to gold nanoparticles. The parameters used were $c_0 = 1.0 \times 10^{15}$ N/m and $\sigma = 0.555$ N/m, with the primary particle radius $r_1 = 2.1 \times 10^{-8}$ nm taken from experiment (cf. Ref. 22). The distributions are shown for two values of the kinetic parameter $f$, at times $t = 0.1, 1.0, 5.0,$ and $10.0$ s. For $f = 0.5$, $t_{sat} = 7.5$ s and $d_{full} = 0.624 \pm 0.02$ $\mu$m. For $f = 10^{-4}$, again $t_{sat} = 7.5$ s, but $d_{full} = 1.973 \pm 0.035$ $\mu$m, in agreement with experiment. See text for an explanation of the units used on the vertical axes.

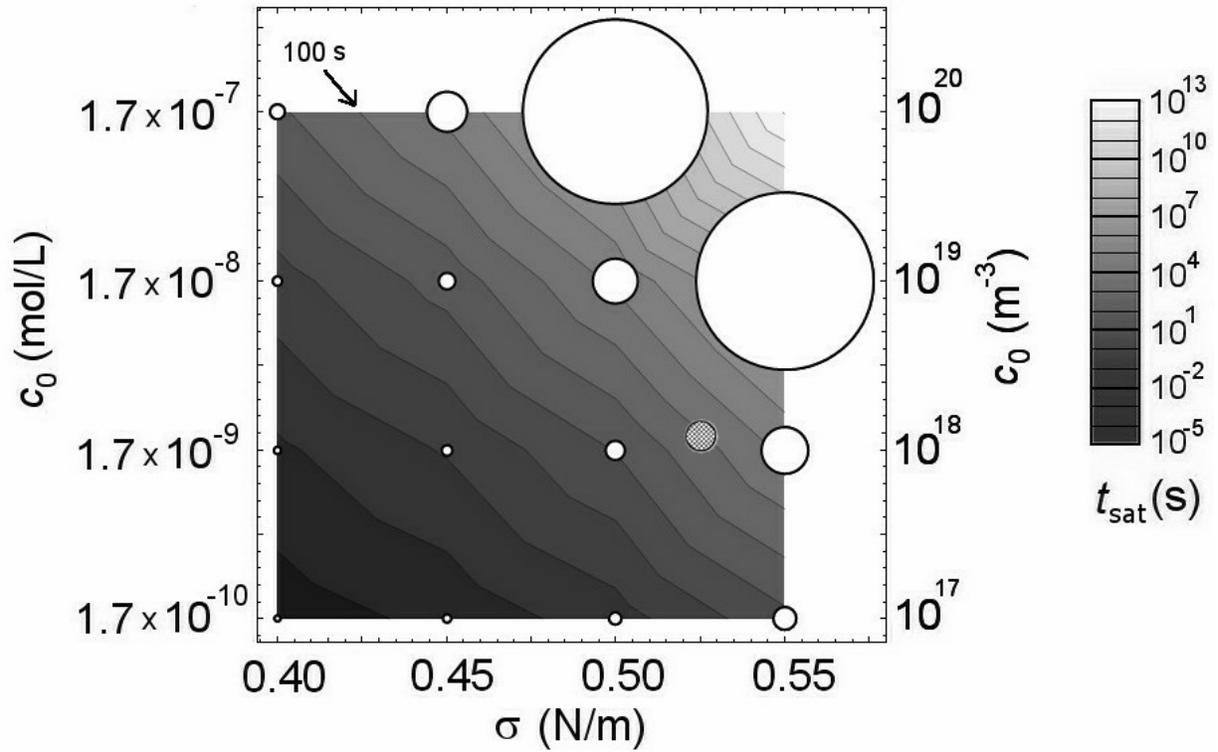

**Figure 6.** Contour plot of the saturation time for simulated Ag secondary particle growth, using kinetic parameter $f = 0.5$, as a function of the surface tension and the equilibrium concentration $c_0$. The experimental time scale of the $T = 60 C$ case in aqueous solution can be fit by any pair of parameter values $(\sigma, c_0)$ drawn from the $t = 100\,\text{s}$ contour. Our best estimate, based on a surface tension value from the literature,[35] $\sigma = 0.525\,\text{N/m}$, is $c_0 = 1.15 \times 10^{18}\,\text{m}^{-3}$. The circles represent the average particle diameter at saturation on the regular grid of $(\sigma, c_0)$ values shown, as well as for the best-fit parameters (shaded circle). The scale is such that the shaded circle corresponds to diameter $d_{\text{full}} = 0.349\,\mu\text{m}$. With kinetic parameter $f = 5.0 \times 10^{-4}$, the diameter for the best-fit parameters is $d_{\text{full}} = 0.970\,\mu\text{m}$, in agreement with the experimental value $d_{\text{expt}} = 0.967\,\mu\text{m}$.

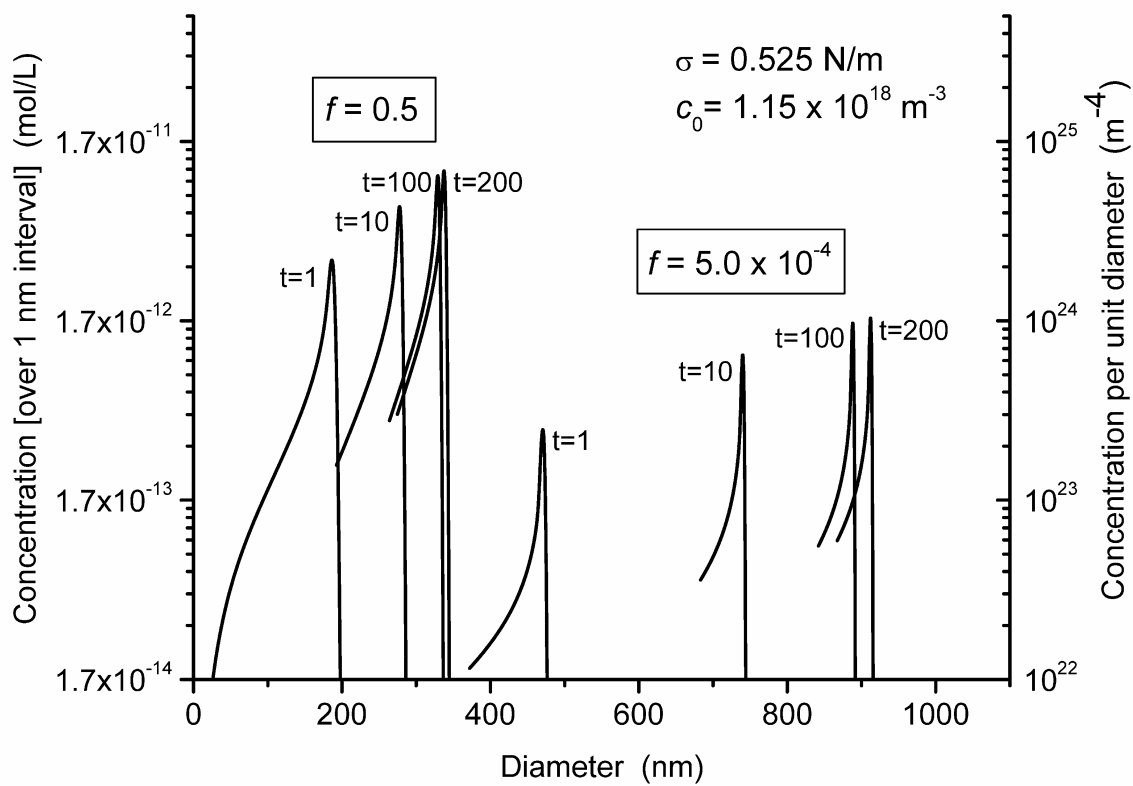

**Figure 7.** Size distribution of Ag nanoparticles vs. time, for best-fit parameter values $\sigma = 0.525$ N/m and $c_0 = 1.15 \times 10^{18}$ m$^{-3}$, corresponding to the shaded circle in Fig. 6. Distributions are shown for two values of the kinetic parameter, $f = 0.5$ and $f = 5 \times 10^{-4}$. With $\sigma$ and $c_0$ chosen to match the experimental time scale, the value of the kinetic parameter required to produce agreement with the experimentally measured Ag particle size is close to that found in Fig. 5 for Au nanoparticles.

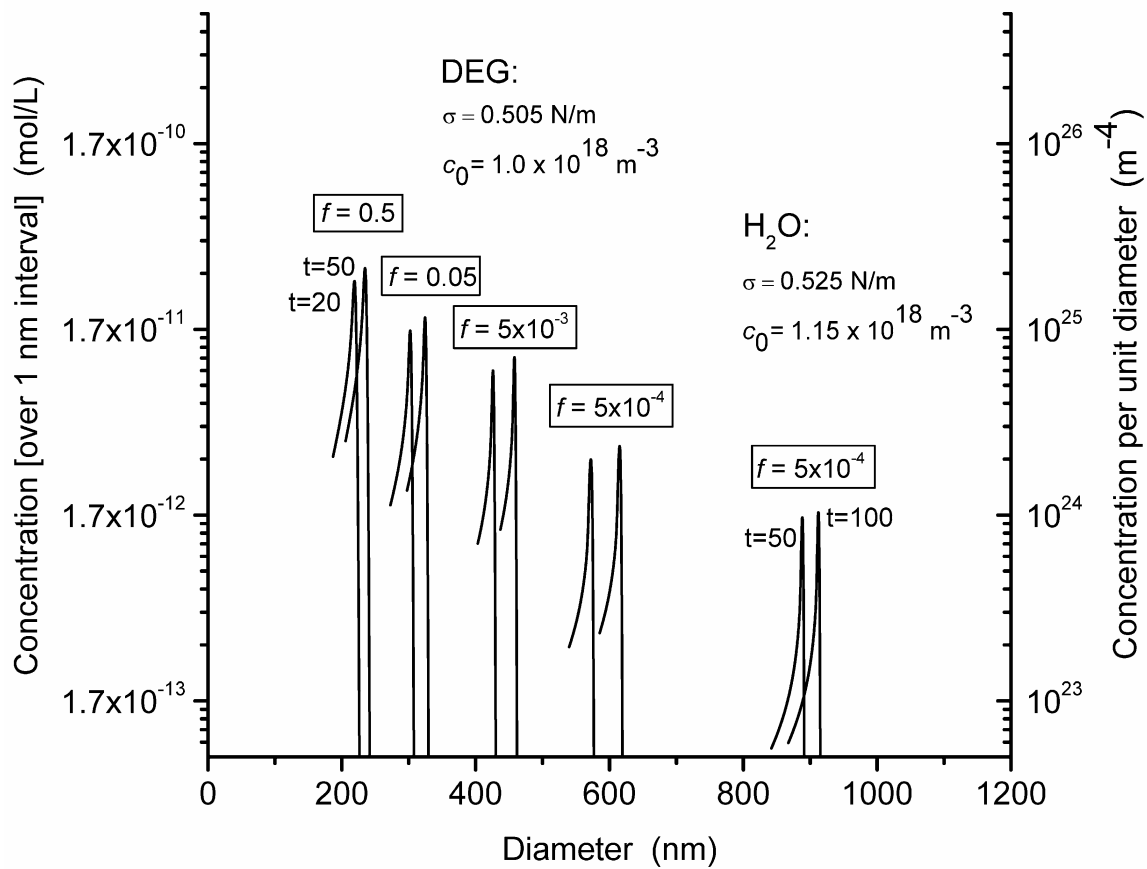

**Figure 8.** Simulated particle size distributions for Ag nanoparticles in aqueous solvent (labeled H$_2$O) and in diethylene glycol (labeled DEG). The simulations were both run at $T$ = 60 C, using best-fit parameter values $\sigma = 0.525$ N/m and $c_0 = 1.15 \times 10^{18}$ m$^{-3}$, and kinetic parameter $f = 5 \times 10^{-4}$.

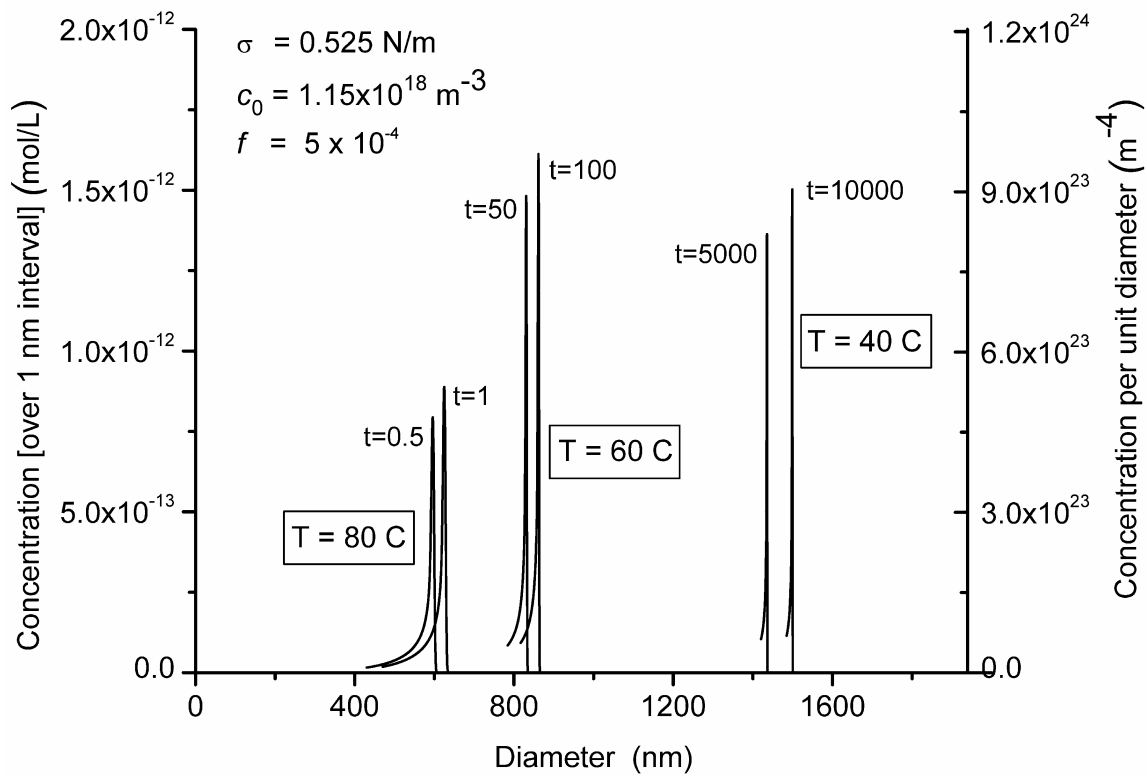

**Figure 9.** Simulated particle size distributions leading up to saturation, for Ag nanoparticles at the three temperatures $T$ = 40, 60, and 80 C. In each case the best-fit parameter values $\sigma = 0.525$ N/m and $c_0 = 1.15 \times 10^{18}$ m$^{-3}$ were used, with kinetic parameter $f = 5 \times 10^{-4}$.